\documentstyle[twocolumn,aps,prc,pstricks,psfig]{revtex}
			\begin{document}
\preprint{LBL-38379R}
\draft

\title{Amplification of pionic instabilities in high-energy collisions?}

\author{J\o rgen Randrup}

\address{Nuclear Science Division, Lawrence Berkeley Laboratory,
University of California, Berkeley, California 94720}

\date{May 1, 1996}

\maketitle

\begin{abstract}
Considering a variety of dynamical scenarios within the linear $\sigma$ model,
we examine the conditions for pionic modes to be amplified
as the chiral field relaxes towards the normal vacuum
following the transient restoration of approximate chiral symmetry
in a high-energy collision.
While Bjorken-type longitudinal expansions appear to be insufficient
for amplification for occur,
analoguos expansions in two or three dimensions
may enhance the low-energy power spectrum by up to an order of magnitude.
\end{abstract}

\pacs{PACS numbers: 25.75.+r,11.30.Rd,11.30.Qc,12.38.Mh}
\narrowtext


The possibility of forming disoriented chiral condensates ({\sl DCC})
in high-energy collisions has generated considerable research activity
in recent years
(for a recent review, see ref.\ \cite{Rajagopal:QGP2}).
The basic premise is that such collisions produce extended hot regions
within which approximate chiral symmetry is temporarily restored.
The subsequent non-equilibrium relaxation
may then lead to the formation of coherent sources of low-energy pions
and associated anomalous multiplicity distributions
\cite{Anselm,Blaizot:PRD46,Rajagopal:NPB399}.
Important insights into the dynamics of {\sl DCC} formation
have been gained with analytical approaches
\cite{Boyanovsky:PRD48,Blaizot:PRD50,MM,Csernai}
and, in the most elaborate studies,
the evolution of the chiral degrees of freedom has been simulated numerically
\cite{Rajagopal:NPB404,%
GGP93,Huang:PRD49,GM,Cooper,Asakawa,Boyanovsky:PRD51,Kluger}.
We present here a simple and instructive framework within which
we analyze the conditions for the occurrence of the {\sl DCC} phenomenon.

Most dynamical studies of disoriented chiral condensates
have been based on the linear $\sigma$ model
in which the chiral degrees of freedom are described by the real $O(4)$ field
$\mbox{\boldmath $\phi$}(\mbox{\boldmath $r$},t)
=(\sigma,\mbox{\boldmath $\pi$})$
having the equation of motion
\begin{equation}
\label{EoM}
\left[\Box+\lambda(\phi^2-v^2)\right]\mbox{\boldmath $\phi$}
=H\hat{\mbox{\boldmath $\sigma$}}\ .
\end{equation}
The three parameters in the model can be fixed by specifying
the pion decay constant, $f_\pi=92\ {\rm MeV}$, 
and the meson masses, $m_\pi=138\ {\rm MeV}$ and $m_\sigma=600\ {\rm MeV}$,
leading to the values
$\lambda=(m_\sigma^2-m_\pi^2)/2f_\pi^2=20.14$,
$v=[(m_\sigma^2-3m_\pi^2)/(m_\sigma^2-m_\pi^2)]^{1/2}f_\pi=86.71\ {\rm MeV}$,
and $H=m_\pi^2 f_\pi= (120.55\ {\rm MeV})^3$,
with $\hbar,c$=1 \cite{JR:dccT}.

As is apparent from eq.\ (\ref{EoM}),
the vacuum configuration is aligned with the $\sigma$ direction,
$\mbox{\boldmath $\phi$}_{\rm vac}=(f_\pi,\mbox{\boldmath $0$})$,
and at low temperature the fluctuations
represent nearly free $\sigma$ and $\pi$ mesons.
In the other extreme, at temperatures well above $v$,
the field fluctuations are centered near zero
and approximate $O(4)$ symmetry prevails.

In order to analyze the situation,
it is instructive to decompose the chiral field,
$\mbox{\boldmath $\phi$}(\mbox{\boldmath $r$},t)
=\underline{\mbox{\boldmath $\phi$}}(t)
+\delta\mbox{\boldmath $\phi$}(\mbox{\boldmath $r$},t)$.
The first term, $\underline{\mbox{\boldmath $\phi$}}$,
is the average over a suitable region in space
(of dimensions larger than the correlation length)
and can be identified with the (local) order parameter,
while the fluctuations, $\delta\mbox{\boldmath $\phi$}(\mbox{\boldmath $r$})$,
may be considered as elementary quasi-particle excitations.
We may simplify the discussion without affecting the conclusions
by assuming that the order parameter is always fully aligned,
$\underline{\mbox{\boldmath $\phi$}}=(\sigma_0,\mbox{\boldmath $0$})$.
The fluctuations are then
$\delta\mbox{\boldmath $\phi$}(\mbox{\boldmath $r$},t)
=(\delta\sigma,\delta\mbox{\boldmath $\pi$})$.

Applying a Hartree-type factorization \cite{Boyanovsky:PRD51,JR:dccT},
we may replace the full equations of motion (\ref{EoM})
by a set of self-consistent approximate equations,
\begin{eqnarray}\label{EoM0}
&~&\left[ \Box + \mu_0^2 \right] \sigma_0 = H\ ,\\ 
\label{EoMp}
&~&\left[ \Box + \mu_\sigma^2 \right] \delta\sigma = 0\ ,\\ 
\label{EoMt}
&~&\left[ \Box + \mu_\pi^2 \right] \delta\mbox{\boldmath $\pi$}
= \mbox{\boldmath $0$}\ ,
\end{eqnarray}
where the associated effective masses are given by
\begin{eqnarray}\label{mu0}
\mu_0^2 &=& \lambda
[\phantom{3}\sigma_0^2\ +<\delta\phi^2> +\ 2<\delta\sigma^2>-\ v^2]\ ,\\
\label{mup}
\mu_\sigma^2 &=& \lambda
[{3}\sigma_0^2\ +<\delta\phi^2> +\ 2<\delta\sigma^2> -\ v^2]\ ,\\
\label{mut}
\mu_\pi^2 &=& \lambda
[\phantom{3}\sigma_0^2\ +<\delta\phi^2> +\ 2<\delta\pi_i^2> -\ v^2]\ ,
\end{eqnarray}
with $\delta\pi_i$ being the particular cartesian component
of $\delta\mbox{\boldmath $\pi$}$.
Thus the field fluctuations provide an additional stiffnes
resisting the growth of the order parameter $\sigma_0$.
The effective masses are degenerate for $\sigma_0$=0
and they vanish at the temperature $T_c=\sqrt{2}v$,
in the mean-field treatment.
Moreover,
we always have $\mu_0^2\leq\mu_\pi^2\leq\mu_\sigma^2$.

Eqs.\ (\ref{EoM0}) and (\ref{mu0})
were first derived in ref.\ \cite{Baym:PRD15}.
The term $<\delta\phi^2>$ in eqs.\ (\ref{mu0}-\ref{mut})
is the sum of the field fluctuations in each of the $N$=4 chiral directions
and is of leading order in $1/N$.
These are the `direct' terms that
have been included in previous {\sl DCC} treatments or discussions
in terms of effective masses
\cite{Boyanovsky:PRD48,Rajagopal:NPB404,GGP93,Boyanovsky:PRD51,Kluger}.
The next term in eqs.\ (\ref{mu0}-\ref{mut})
is twice the fluctuation along the particular
direction considered (either parallel or perpendicular to the order parameter)
and arises from the `exchange' terms.
A recent analysis of the statistical properties of the linear $\sigma$ model
\cite{JR:dccT} suggests that their effect is significant ($\sim2/N$=50\%)
and that their inclusion leads to a quite good approximation
to the full eq.\ (\ref{EoM}).

The approximate equations (\ref{EoM0}-\ref{mut}) provide a convenient framework
for developing a qualitative understanding of the dynamics
generated by the full equation (\ref{EoM}).
Imagine that the system is initially created in thermal equilibrium
at a temperature $T_0$ well above $T_c$.
The field fluctuations are then sufficiently large to ensure $\mu^2$$>$0
in all three eqs.\ (\ref{mu0}-\ref{mut}).
The system is expected to experience a cooling
resulting from expansion and radiation,
so the fluctuations decrease in the course of time.
This reduces $\mu^2$ which allows the order parameter to grow larger,
thus counteracting the decrease of the effective masses.
The resulting behavior of $\mu^2$ is then a delicate balance:
for slow cooling the induced growth of $\sigma_0$ is approximately adiabatic
and the system relaxes towards the vacuum through metastable configurations;
however, if the fluctuations diminish rapidly
a compensating growth of the order parameter
can no longer occur quickly enough
and one or more of the effective masses may turn imaginary, $\mu^2$$<$0,
indicating that the system has entered a regime
exhibiting exponential growth of some modes.

\begin{figure}
\vfill	
~\hfill	
\psfig{figure=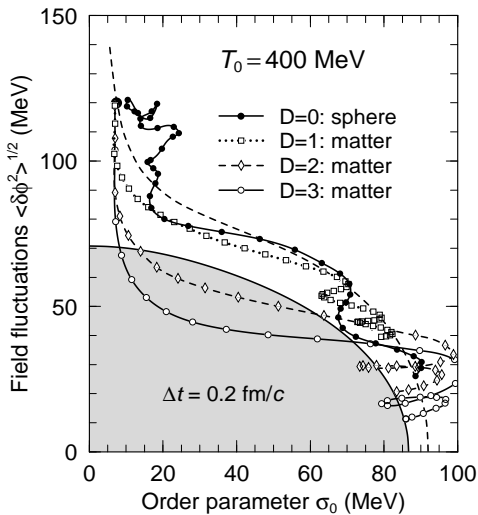,width=3.0in}
\caption{Dynamical trajectories.}
\label{f:T400}
The combined dynamical evolution
of the order parameter $\sigma_0=<\sigma>$
and the field fluctuations $<\delta\phi^2>^{1/2}$
(sampled for $r<$2.5 fm in the case of $D$=0).
The dashed curve connects the equilibria
from $T$=0 to above 500 MeV
and the unstable region within which $\mu_\pi^2<0$ is shown
by the shaded region.
Each system has been prepared in thermal equilibrium
at $T_0=400$ MeV,
using a periodic box (20 fm side length).
The irregular solid trajectory ($D$=0)
was obtained by solving the standard eq.\ (\ref{EoM})
after applying a spherical Saxon-Woods modulation factor
(5 fm radius and 0.5 fm width) to the hot matter,
thereby producing a hot sphere embedded in vacuum \cite{Paulo}.
The other three trajectories have been obtained without a spatial modulation
but with the term $-(D/t)\partial_t\mbox{\boldmath $\phi$}$ added
in the equation of motion
to emulate uniform expansions in $D$=1,2,3 dimensions.
The marks along the trajectories are positioned
at time intervals of $\Delta t=0.2\ {\rm fm}/c$.
\end{figure}

Fig.\ \ref{f:T400} shows the equilibria and the unstable region.
As $T$ is increased,
the fluctuations grow steadily and
the equilibrium value of $\sigma_0$ decreases from $f_\pi$.
The most rapid change occurs at $T\approx$220 MeV,
above which $\sigma_0$ tends to zero.
The effective pion mass $\mu_\pi$ increases monotonically with $T$
from its free value $m_\pi$ towards $\approx 1.6T$ for $T\gg T_c$,
while $\mu_\sigma$ first decreases,
then displays a minimum at $T\approx240\ {\rm MeV}$,
and finally becomes degenerate with $\mu_\pi$.
The border of the unstable region intersects the $\sigma$ axis at $f_\pi$
and extends up to $T_c/\sqrt{3}$ at $\sigma_0$=0.

Also shown in fig.\ \ref{f:T400}
is the dynamical trajectory of the central part
of a Ni-sized spherical source prepared at $T_0$=400 MeV.
The system keeps away from the unstable regime,
exhibiting an approximately adiabatic evolution.
(The windings reflect the damped oscillations
of $\sigma_0$ in the relaxing effective potential,
in the familiar manner \cite{Huang:PRD49,GM}.)
This behavior is rather robust,
as it occurs for a wide range of initial temperatures
and for rod or slab geometries as well.
It thus appears that initially static field configurations
in local equilibrium do not develop any instabilities
during their subsequent expansion.

However,
it is expected that the early parton dynamics
causes the chiral field to be in a state of rapid expansion.
The subsequent evolution may then lead to
a supercooled configuration situated inside the unstable region,
thus effectively producing a ``quench''.
A number of quenched scenarios have been considered
\cite{Rajagopal:NPB404,Huang:PRD49,GM,Cooper,Asakawa,Boyanovsky:PRD51,Kluger}
but they were imposed by {\em fiat},
thereby reducing the predictive power of the dynamcial calculations
(essentially any degree of magnification can be achieved
by suitable adjustment of the initial conditions).
It is our aim here to reduce the degree of arbitrariness
by elucidating under which conditions a quench-like early scenario
may develop dynamically from plausible initial configurations.

Simple Bjorken-like pictures
can be invoked to emulate expanding scenarios in $D$ dimensions,
either longitudinal ($D$=1) \cite{Huang:PRD49,Cooper,Kluger},
transverse ($D$=2), or isotropic ($D$=3) \cite{GM,Lampert}.
We have considered such scenarios in the following approximate manner.
The essential effect of the expansion is the appearance of
an additional term on the right-hand side of eq.\ (\ref{EoM}),
$-(D/t)\partial_t\mbox{\boldmath $\phi$}$,
where the time variable should now be reinterpreted as the elapsed proper time
in a comoving frame (starting at $t_0=1\ {\rm fm}/c$, usually).
The corresponding Lorentz transformation of the (scaled) spatial coordinates
is less essential for our present discussion
and has therefore been ignored.

\begin{figure}
\vfill	
~\hfill	
\psfig{figure=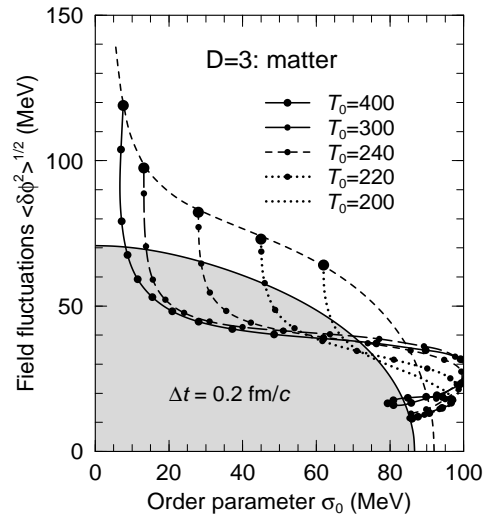,width=3.0in}
\caption{Idealized isotropic expansions.}
\label{f:D3}
Similar to fig.\ \ref{f:T400},
but for idealized isotropic expansions
using $D$=3 in the modified equation of motion.
\end{figure}

The additional term in the equation of motion,
$-(D/t)\partial_t\mbox{\boldmath $\phi$}$,
acts as a time-dependent damping,
its form being akin to the Rayleigh dissipation function in classical mechanics,
and it is significantly more effective in reducing the fluctuations
than self-generated expansions.
In order to examine its effect,
we ignore the spatial geometry
and consider a macroscopically uniform configuration
within a large box.
The discussion is then simplified and
the resulting scenarios can be regarded as
idealized representations of chiral matter subjected to 
an externally prescribed cooling rate,
and so the results will have a corresponding general applicability.

Figure \ref{f:T400} includes dynamical trajectories
obtained in this manner for $D$=1,2,3.
The effect increases with $D$,
since the dimensionality of the expansion
effectively acts as the strength of the damping term.
The isotropic expansion leads to
a significant incursion into the unstable region,
while the longitudinal one is too slow for that.

\begin{figure}
\hspace{-3.1in}
\psfig{figure=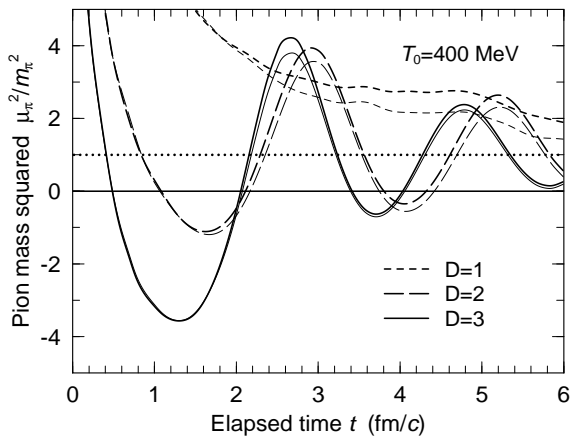,width=3.0in}
\caption{Time evolution of the pion mass.}
\label{f:mu2}
The time evolution of $\mu_\pi^2$,
the square of the effective pion mass, eq.\ (\ref{mut}),
for idealized $D$-dimensional expansions
starting from thermal equilibrium at $T_0$=400 MeV.
The thin curves show the corresponding evolution of $\mu_0^2$
(recall $\mu_0^2\leq\mu_\pi^2\leq\mu_\sigma^2$).
All curves approach the free pion mass $m_\pi$ (dotted line)
in the course of time.
\end{figure}

\begin{table}[h]
\caption{Amplification coefficients/correlation lengths.}
\label{table}
\begin{tabular}{|c|ccc|}
$T_0$ (MeV)	& $D=1$		& $D=2$		& $D=3$		\\
\hline	
200		& 0.00	/1.4	& 0.02	/1.8	& 0.11	/2.0	\\
220		& 0.00	/1.3	& 0.50	/1.9	& 0.55	/2.5	\\
240		& 0.01	/1.3	& 1.20	/2.0	& 1.19	/2.7	\\
300		& 0.00	/0.9	& 1.84	/1.7	& 2.06	/2.7	\\
400		& 0.00	/0.6	& 1.67	/1.3	& 2.49	/2.1	\\
500		& 0.00	/0.5	& 1.31	/1.1	& 2.61	/1.6	
\end{tabular}	
The enhancement factor $G^\pi_{k=0}$ in macroscopically uniform chiral matter
and the resulting correlation length,
expressed as the
full width at half maximum (in fm) of the spatial two-point function
$\prec\delta\mbox{\boldmath $\pi$}(x)\cdot
\delta\mbox{\boldmath $\pi$}(y)\succ$,
for pions emerging after idealized expansions with $D$=1,2,3
starting from thermal equilibrium at $T_0$.
\end{table}

Figure \ref{f:D3} shows trajectories for $D$=3
starting from various temperatures.
If the initial temperature is lower than 200 MeV or so,
the initial value of $\sigma_0$ is already fairly large (over 60 MeV)
and the dynamical trajectories will miss the unstable region.
A wide range of higher temperatures lead into the unstable region,
provided the supercooling occurs sufficiently fast.
Ultimately, at very high temperatures (above those shown)
the system will again stay stable throughout,
because it takes so long to reduce the fluctuations down to critical size
that the order parameter has meanwhile had time to start its growth.

In order to quantify the analysis,
it is useful to consider the time evolution of the effective masses.
Since $\mu_\sigma^2>\mu_\pi^2$
we concentrate on $\mu_\pi^2$ which is illustrated in fig.\ \ref{f:mu2}.
It is noteworthy that $\mu_0^2\approx\mu_\pi^2$ throughout the evolution.
(Early on $\sigma_0$ is small so $O(4)$ symmetry holds approximately,
and later on the fluctuations are less important than $\sigma_0$.)
The amplification of the lowest pionic modes
is then practically identical to that of the order parameter itself.
This simple feature makes it an easier task to analyze
more complicated scenarios.
As was already evident from figs.\ 1-2,
multiple incursions into the unstable region are possible,
especially for large initial cooling rates
(since a rapid quench brings $\sigma_0$ into larger
oscillations around $f_\pi$),
but the first one is always dominant.

It is convenient to express the resulting enhancement of a mode
in terms of its amplification coefficient \cite{HH},
\begin{equation}\label{G0}
G^\pi_k\ \equiv\	
\int_{\omega_k^2<0}dt\ \sqrt{-\omega_k(t)^2}\ ,
\end{equation}
where the frequency is given by the dispersion relation,
$\omega_k^2=k^2+\mu_\pi^2$,
using the time-dependent effective pion mass (\ref{mut}).
[The dispersion relations for the quasiparticles
are in principle affected by the presence of the expansion term,
as is ordinarily the case for a damped oscillator.
This effect may be significant for damping rates large enough
to bring the system into the unstable region,
but has been ignored in the present study.]
The quantity $\exp(G^\pi_k)$ expresses approximately the factor by which
the amplitude of pions having the wave number $k$ has been magnified
due to the incursion(s) into the unstable regime.
Since largest magnification occurs for $k$=0
(a finite $k$ adds a positive amount to $\omega_k^2$),
we concentrate on this quantity as an upper bound.
Moreover, we note that the minimum in $\mu_\pi^2(t)$ (see fig.\ \ref{f:mu2})
determines the maximum wave number for which magnification occurs.
Amplification coefficients obtained for various idealized expansion
scenarios are shown in table \ref{table},
together with the resulting width of the pion correlation function.

The purely longitudinal expansions ($D$=1) miss the unstable region,
except for insignificant incursions near $T$=240 MeV,
while a significant degree of magnification occurs for
the transverse and isotropic expansions,
amounting to over a factor of ten in the most favorable cases.
It should be recalled that $G^\pi_0$ applies to $k$=0 only,
so it provides an upper bound on the enhancement.

\begin{figure}
~\vspace{0.2in}
\hspace{-3.2in}
\psfig{figure=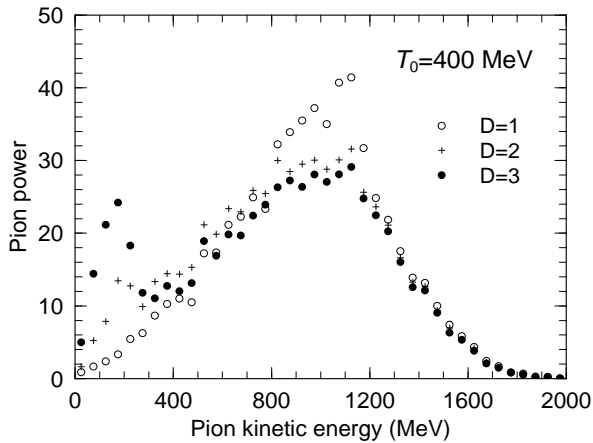,width=3.0in}
\caption{Pion power spectrum.}
\label{f:power}
The relative power spectrum of the pions,
$\sim\omega_k^2\pi_k^2$,
where $\mbox{\boldmath $\pi$}_k$ is the Fourier amplitude of the pion field,
plotted as a function of the pion kinetic energy $\omega_k$-$m_\pi$.
The extraction is made at large times
when the asymptotic scenario of free evolution has been reached.
The plots are based on samples of 20 field configurations
prepared at $T_0$=400 MeV
and subjected to idealized expansions
with either $D$=1 (open dots), $D$=2 (crosses), or $D$=3 (solid dots).
The irregularities are primarily due to the shell structure
in the level density of the cube.
\end{figure}

Figure \ref{f:power} gives an impression of the net effect
on the power spectrum of the emerging free pions.
As expected,
the transient instabilities present for $D$=2,3
lead to significant enhancements of the power carried off
by soft pions.
The effect amounts to about an order of magnitude for $D$=3
(relative to the smooth spectrum obtained for $D$=1),
in accordance with the amplification coefficient given in Table \ref{table}.
Although these results were calculated for idealized expansion scenarios,
they do support the suggestion that such enhancements
may provide an observable signal of {\sl DCC} formation \cite{Gavin:QM95}.

It should be emphasized
that the dynamical evolution of the chiral field
was obtained by solving the full equation of motion (\ref{EoM}).
The simple Hartree approximation (\ref{EoM0}-\ref{mut})
has been invoked only as a convenient framework for the discussion
since it brings out so clearly the conditions for amplification.

Our analysis shows that the occurrence of instabilities,
and the associated amplification of pionic modes,
depends sensitively on the cooling rate,
which in turn is intimately related to the character of the expansion.
Our idealized scenario for $D$=3
corresponds closely to the isotropic expansion
considered in refs.\ \cite{GM,Lampert}
and our results corroborate the conclusion in \cite{GM}
that such a scenario leads to amplification.
Furthermore,
our analysis suggest that a longitudinal expansion alone
is insufficient to cause a quench,
if the initial fluctuations are of thermal magnitude.
This is consistent with what was found
in refs.\ \cite{Huang:PRD49,Kluger}
for effectively one-dimensional expansions.

This qualitative sensitivity to the collision dynamics
highlights the importance of employing realistic initial conditions
for the dynamical simulations of {\sl DCC} formation.
Ultimately,
the appropriate initial field configurations
must be calculated on the basis of the early partonic evolution,
a task which is thus crucial for our ability to assess the prospects
of forming disoriented chiral condensates in high-energy collisions.

~\\ \noindent
The author wishes to acknowledge helpful discussions with
S. Gavin, R. Vogt, and X.-N. Wang.
This work was supported by the Director, Office of Energy Research,
Office of High Energy and Nuclear Physics,
Nuclear Physics Division of the U.S. Department of Energy
under Contract No.\ DE-AC03-76SF00098.\\



\vfill\noindent
Preprint LBL-38379R: \hfill {\sl Physical Review Letters}\\

			\end{document}